\documentclass[final,1p,times,twocolumn]{elsarticle}

\usepackage{graphicx}

\usepackage{amssymb}

\journal{Comptes Rendus Physique}

\begin{document}

\begin{frontmatter}



\title{Properties of Ferromagnetic Superconductors}


\author[cea]{Dai Aoki}
\author[kit]{Fr\'{e}d\'{e}ric Hardy}
\author[osaka,cea]{Atsushi Miyake}
\author[cea]{Valentin Taufour}
\author[jaea,cea]{Tatsuma D. Matsuda}
\author[cea]{Jacques Flouquet}

\address[cea]{SPSMS, UMR-E CEA / UJF-Grenoble 1, INAC, 38054 Grenoble, France}
\address[kit]{Karlsruher Institut f\"{u}r Technologie, Institut f\"{u}r Festk\"{o}rperphysik, 76021 Karlsruhe, Germany}
\address[osaka]{KYOKUGEN, Osaka University, Toyonaka, Osaka 560-8531, Japan}
\address[jaea]{Advanced Science Research Center, Japan Atomic Energy Agency, Tokai, Ibaraki 319-1195, Japan}

\begin{abstract}
Thanks to the discovery in the last decade of three uranium ferromagnetic superconductors,
UGe$_2$, URhGe and UCoGe,
the fascinating aspects of the interplay between the triplet state of Cooper pairing and ferromagnetism have emerged.
Furthermore, as the ferromagnetic properties in the normal state are quite different 
with respect to the proximity of the ferromagnetic--paramagnetic instabilities,
the feedback with the coexistence of superconductivity gives rise to quite different boundaries 
in pressure and magnetic field.
Special attention is given on the location of the materials with respect to the tricriticality and 
on the reinforcement of SC in a transverse field response with respect to the direction of the FM sublattice magnetization.
The other facts of the interplay between FM and SC is briefly mentioned.
\end{abstract}
\end{frontmatter}


\section{Introduction}
The discovery of superconductivity (SC) in the ferromagnet UGe$_2$ has opened a new chapter in the exotic domain of unconventional superconductivity~\cite{Sax00}. 
The trend is that the ferromagnetic (FM) interaction between highly renormalized quasiparticles is the source of SC pairing. 
In the three Ising ferromagnetic superconductors UGe$_2$, URhGe~\cite{Aok01} and UCoGe~\cite{Huy07}, it appears that the Cooper pairs condense in the equal spin pairing state (ESP) with $\uparrow\uparrow$ and $\downarrow\downarrow$ spin carriers. 
In this review, we give a schematic view of the phenomenon. 
This article is quite complementary to the paper recently published in J. Phys. Soc. Jpn. for the 100 years of superconductivity~\cite{Aok11_JPSJ_review}. 
We focus on temperature ($T$), pressure ($P$) and magnetic field ($H$) phase diagrams,
in particular on the precise location of the FM and FM+SC phases, and the PM (paramagnetic) and PM+SC boundaries. 
In these compounds, the occurrence of SC is strongly related to the effective mass enhancement associated with the ferromagnetic instability
which occurs in UCoGe at the critical point ($P_{\rm c}$,T=0) where FM is collapsed,
while in UGe$_2$ two distinct ferromagnetic phases FM1 and FM2 are separated by $P_{\rm x}$.
The new feature of these Ising ferromagnets is that
the field response of the FM--PM instability is quite anisotropic between
$H \parallel M_0$ and $H \perp M_0$, where $M_0$ is the sublattice magnetization.
Furthermore, when the system moves towards $P_{\rm c}$, the FM-PM transition line, $T_{\rm Curie}(P)$, becomes first order and the occurrence of tricriticality at ($T_{\rm TCP}$,$P_{\rm TCP}$) leads to the existence of a field induced FM phase which extends beyond $P_{\rm c}$ till a quantum critical end point (QCEP) at ($P_{\rm QCEP}$,$H_{\rm QCEP}$).

In this article, first we briefly describe the features of itinerant ferromagnetism and SC pairing mediated by ferromagnetic fluctuations 
and then we summarize the normal-state properties of UGe$_2$, URhGe and UCoGe and comment on their influence on the appearance of SC at zero field. 
We discuss the occurrence of tricriticality in UGe$_2$ and the $H$ reinforced/reentrant SC for $H\perp M_0$ in URhGe and UCoGe in the context of their longitudinal and transverse field responses~\cite{Lev05,Aok09_UCoGe,Aok11_ICHE}. In conclusion, we give a short list of other aspects of the interplay between FM and SC.

\section{Ferromagnetism in itinerant electronic system}
\begin{figure}[tbh]
\begin{center}
\includegraphics[width=0.5 \hsize,clip]{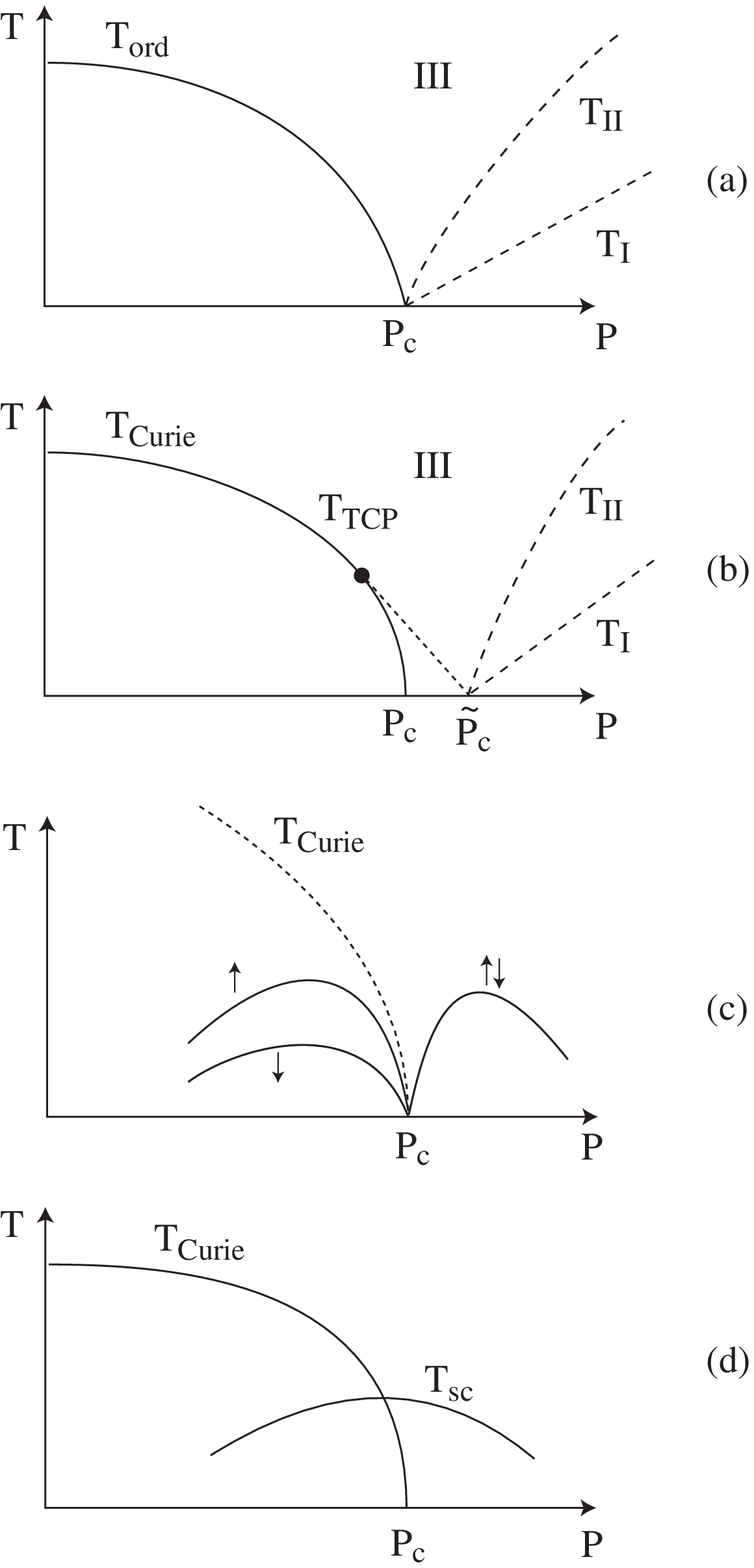}
\end{center}
\caption{(a)Magnetic phase diagram from the SCR theory. Below $T_{\rm I}$, the Fermi liquid properties are observed. For a first order transition, $P_{\rm c}$ can reach $\tilde{P}_{\rm c}$ if $\Delta M_0$ is close to zero (b). (c)Superconducting phase diagram near the ferromagnetic instability, from Fay and Appel. (c)Phase diagram of UCoGe.}
\label{fig:phase}
\end{figure}

A major breakthrough in the understanding of FM in itinerant systems appeared in 1985 with Moriya's self-consistent renormalization (SCR) theory of spin fluctuations in a Hubbard scheme. 
In this model, the Fermi liquid regime, 
which is characterized by a specific-heat linear term $\gamma$ and 
a $T^2$ dependent resistivity below a temperature $T_{\rm I}$, collapses 
when the transition from a long range FM order to a PM ground state occurs at a characteristic pressure $P_{\rm c}$.
On the other hand, the non-Fermi liquid (NFL) regime between $T_{\rm I}$ and $T_{\rm II}$ expands before recovering a high temperature domain $T_{\rm III}$ (see Fig.~\ref{fig:phase}(a))~\cite{Mil93,Flo06_review}. 
Pressure often tunes the system from FM to PM since it increases the electronic bandwidth $W$ and thus decreases the density of states $N(\varepsilon_{\rm F})$.
Below $P_{\rm c}$, $U N(\varepsilon_{\rm F})$ is larger than $1$ while above $P_{\rm c}$, $U N(\varepsilon_{\rm F})$ is smaller than $1$,
where $U$ is the onsite Coulomb repulsion and $N(\varepsilon_{\rm F}^{})$ is the electronic density of state at the Fermi level.
Table~\ref{tab:1} summarizes the expected $P$ dependence of $T_{\rm I}$, $T_{\rm II}$ and $T_{\rm Curie}$ together with the variation of $T_{\rm Curie}$ with $M_{0}$.

\begin{table}[tbhp]
\caption{Pressure dependence of $T_{\rm I}$, $T_{\rm II}$, $T_{\rm Curie}$ and the relation between the sublattice magnetization $M_0$ and $T_{\rm Curie}$ for 3D FM systems.}
\begin{center}
\begin{tabular}{cccccc}
\hline
	& $T_{\rm I}$			& $T_{\rm II}$			& $T_{\rm II}/T_{\rm I}$	& $T_{\rm Curie}$		& $T_{\rm Curie}(M_0)$	\\
\hline
FM	& $(P-P_{\rm c})^{3/2}$	& $(P-P_{\rm c})^{3/4}$	& $(P-P_{\rm c})^{-3/4}$	& $(P-P_{\rm c})^{3/4}$	& $M_0{}^{3/2}$			\\
\hline
\end{tabular}
\end{center}
\label{tab:1}
\end{table}%
\begin{table}[tbhp]
\caption{Pressure dependence of $\gamma$, $\chi_{Q=0}$ and $A$ near $P_{\rm c}$ for 3D FM systems.}
\begin{center}
\begin{tabular}{cccc}
\hline
	& $\gamma$					& $\chi_{Q=0}$				& $A$	\\
\hline
FM	& $\log (P-P_{\rm c})$	& $(P-P_{\rm c})^{-1}$	& $(P-P_{\rm c})^{-1}$	\\
\hline
\end{tabular}
\end{center}
\label{tab:2}
\end{table}%
\begin{table}[tbhp]
\caption{Temperature variation of $C/T$, $chi_{Q=0}$ and resistivity for 3D and 2D FM systems.}
\begin{center}
\begin{tabular}{cccc}
\hline
		& $C/T$			& $1/\chi_{Q=0}$	& $\rho\sim T^n$\\
\hline
FM	3D 	& $-\log T$	& $T^{4/3}$	& $T^{5/3}$	\\
FM	2D	& $T^{-1/3}$	& $-T\log T$	& $T^{4/3}$	\\
\hline
\end{tabular}
\end{center}
\label{tab:3}
\end{table}%
Below $T_{\rm I}$, the effective mass $m^\ast$ taken from $\gamma\sim m^\ast k_{\rm F}$ diverges as $\log (P-P_{\rm c})$,
while the uniform susceptibility $\chi(0)$ and the resistivity inelastic term $A$ diverge as $\frac{1}{P-P_{\rm c}}$.
In addition, the temperature variation of $C/T$ and $\chi(0)$ and the exponent $n$ of the resistivity term depend on the magnetic dimensionality as shown in Table~\ref{tab:3}.
However, there is a major difference between FM and antiferromagnetic (AFM) quantum critical points. For the latter no divergence of $m^\ast$ occurs at $P_{\rm c}$ and $\gamma$ varies as $\gamma_0 ( P-P_{\rm c})^{1/2}$. 
The $(P-P_{\rm c})^{1/2}$ singularity leads to a divergence of the Gr\"{u}neisen parameter $\Omega_{\rm e} = - \partial \log \gamma / \partial \log V$ at $P_{\rm c}$. For a second-order phase transition, both the entropy ($S$) and the thermal expansion ($\partial S/\partial P$) collapse at $P_{\rm c}$ (Fig.~\ref{fig:phase}(a)).
 
For FM systems, it is well established both experimentally~\cite{Uhl04} and theoretically~\cite{Bel05} that the divergence of $m^\ast$ at $P_{\rm c}$ is inhibited by the occurrence of a first-order transition at $P_{\rm c}$ (Fig.~\ref{fig:phase}(b)) which is characterized by discontinuities $\Delta M_0$, $\Delta V_0$ in $M_0$ and volume $V_0$, respectively.
As the entropy reaches zero at $P=P_{\rm c}$ according to the Clausius-Clapeyron relation ($d P /d T = \Delta S / \Delta V$),
the initial $(T,P)$ line at very low temperature must be vertical. 
If $\Delta M_0$ is small, the quantum phase transition at $P_{\rm c}$ will only be weakly first-order
and strong fluctuations will persist, being almost like the second order phase transition.
Thus for a strong first order transition (large $\Delta M_0$ and large $\Delta V$)
there is a large difference between $P_{\rm c}$ and $\tilde{P}_{\rm c}$.

The SCR theory was developed for 3D itinerant magnets and extended to the case of heavy fermion systems (HFS) with
the simple idea that the bandwidth $W$ is renormalized to a Kondo energy $k_{\rm B}T_{\rm K}$
characteristic of the strong local nature of the magnetism and its fluctuations~\cite{Tak96}.
This leads to a strongly renormalized band mass $m_{\rm B}$ and a further enhancement of $m^{\ast\ast}$ due to the FM quasiparticle interactions~\cite{Miy08}.
Very often $m_{\rm B}$ and $m^{\ast\ast}$ have comparable amplitudes. Thus, the image of interfering quasiparticles is that of interfering waves with a large diffraction pattern given by the strong local character of the magnetism.

In the case of cerium HFS, the effect of pressure is to switch the system from a magnetically ordered state to a PM ground state.
This is due to the strong $P$ increase of the Kondo energy $k_{\rm B}T_{\rm K}$ in comparison with the indirect intersite coupling, given by the Ruderman, Kittel, Kasuya, Yosida (RKKY) interaction.
Pressure drives the Ce systems from a trivalent configuration (with a 4$f$-shell occupation number $n_f \sim 1$) to a tetravalent configuration  with $n_f \sim 0$. According to the 4$f$ electron-hole symmetry, $n_f$ can vary from $n_f=14$ (Yb$^{2+}$) to $n_f=13$ (Yb$^{3+}$) in ytterbium HFS and magnetic ground states appear under pressure~\cite{Flo06_review}.
For uranium compounds, it is difficult to predict the pressure dependence of $T_{\rm Curie}$ because the fluctuations now occur between the two magnetic configurations U$^{3+}$ and U$^{4+}$.

\section{Cooper pairing and ferromagnetism}
Soon after the elaboration of the BCS theory of $s$-wave superconductivity,~\cite{Bar57} 
the problem of coexistence of SC and FM was discussed by V. Ginzburg. 
He noticed that finding SC in ferromagnets is as probable as finding non-ferromagnetic SC in large magnetic fields~\cite{Gin57}. 
However, the relevance of FM spin fluctuations for SC was pointed out in 1966~\cite{Ber66}.
The existence of an anisotropic BCS state was illustrated by the $p$-wave superfluidity observed in liquid $^3$He~\cite{Nak73,Bri74,Leg75}. 
$p$-wave SC transitions for paramagnon mediated SC in nearly FM systems were first calculated by Layzer and Fay in 1971~\cite{Lay71}. 
However, it is only in 1980 that Fay and Appel published the first paper concerning the variation of $T_{\rm sc}$ through $P_{\rm c}$ in the limited context of the so called equal spin pairing (ESP) state with $\uparrow\uparrow$ and $\downarrow\downarrow$ quasiparticles (Fig.~\ref{fig:phase}(b))~\cite{Fay80}. 
The ESP interaction with $\uparrow\uparrow$ and $\downarrow\downarrow$ components of the triplet channel with an angular momentum $q$ is related to the non interacting Lindhard response of the spin
$\chi_0^{\uparrow}$ and $\chi_0^{\downarrow}$ by the relation:
\begin{equation}
V_{\uparrow\uparrow} = \frac{\Delta^2 \chi_0^{\downarrow}}
                            {1-U^2 \chi_0^{\downarrow}(q)\chi_0^{\uparrow}(q)}
\end{equation}
To mediate SC with a $\uparrow\uparrow$ minority-spin component,a majority-spin component $\downarrow\downarrow$ is required. 
As shown in Fig.~\ref{fig:phase}(b), the $\uparrow\uparrow$ minority-spin carriers first condense in the FM state and SC corresponds to a two-band model. 
In the PM region, both components condense at the same critical temperature. 
However, if FM abruptly disappears through a first-order transition at $P_{\rm c}$ instead of $\tilde{P}_{\rm c}$, it is then clear that the singularity at $P_{\rm c}$ could be suppressed (Fig.~\ref{fig:phase}(b)). 

In the theory of Fay and Appel performed for the second order transition, the superconducting critical temperature $T_{\rm sc}$ is described by
\begin{equation}
T_{\rm sc} = \omega_{\rm c} \exp\left(-\frac{1+\lambda_z}{\lambda_{\Delta}}\right),
\end{equation}
where $\lambda_z$ is the renormalized-mass parameter and $\lambda_{\Delta}$ is the interaction parameter. $\omega_{\rm c}$, which is basically proportional to $T_{\rm I}$, vanishes at $P_{\rm c}$. Close to $P_{\rm c}$, this formula differs from the well-known McMillan-like formula, 
\begin{equation}
T_{\rm sc}  \sim T_0 \exp ( -1/\lambda ),
\end{equation}
with
\begin{equation}
\lambda=\frac{\lambda_{\Delta}}{1+\lambda_z},
\end{equation}
and where $T_0$ is a characteristic cutoff energy. Outside around $P_{\rm c}$, they are basically the same. For URhGe, a simpler expression was chosen,~\cite{Miy08}
\begin{eqnarray}
T_{\rm sc} & \sim& T_0 \exp\left( -\frac{m^\ast}{m^{\ast\ast}} \right) \\
 m^\ast    &=    & m_{\rm B}+m^{\ast\ast},
\end{eqnarray}
where the quasiparticle effective mass $m^\ast$ is the sum of the band mass $m_{\rm B}$ and the correlation mass $m^{\ast\ast}$. Here $1+\lambda_z = m^\ast / m_0$ and $\lambda_{\Delta} = m^{\ast\ast}/m_0$, where $m_0$ is the free electron mass. 
Further calculations of $T_{\rm sc}(P)$ show that
$T_{\rm sc}$ has only weak minima at $P_{\rm c}$~\cite{Rou01}. 
Additional discussions concerning the coexistence of FM and SC can be found in Refs.\cite{Kir01,Kar01,Nev05,Kar08,Sho09}.
Calculations in the PM side of $P_{\rm c}$, for AF and FM interactions, were performed using the Eliashberg formalism for the quasi-2D and 3D cases~\cite{Mon01} with specific applications to cubic and tetragonal symmetries as a function of the electronic or magnetic anisotropy~\cite{Mon04}. 
In general, a spin singlet is favored. 
However, for a triplet state, pairing is only caused by longitudinal fluctuations whereas transverse fluctuations are pair-breaking and impurity scattering is strongly enhanced at $P_{\rm c}$~\cite{Miy02}. 
Thus, it is not surprising that triplet SC in ferromagnets has only been discovered in Ising ferromagnets. 
Finally, the possible SC order parameters in ferromagnetic materials have been classified using general symmetry arguments for cubic and orthorhombic structures~\cite{Sam02,Fom01,Min02}.

\section{Three ferromagnetic superconductors: UGe$_2$, URhGe and UCoGe}
\begin{figure}[tbh]
\begin{center}
\includegraphics[width=0.6 \hsize,clip]{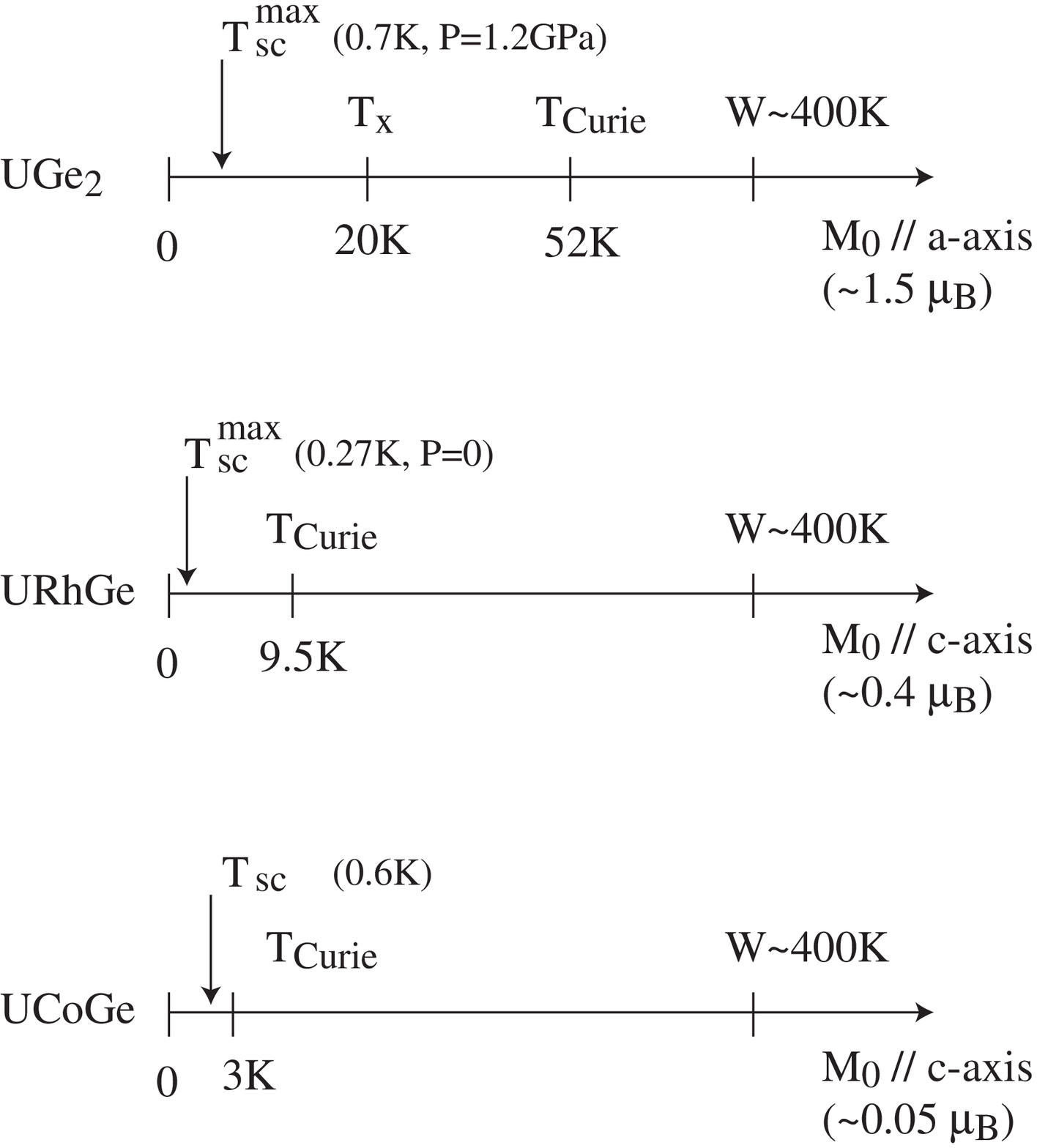}
\end{center}
\caption{Characteristic energy scales of the three ferromagnetic superconductors, UGe$_2$, URhGe and UCoGe.}
\label{fig:energy_scale}
\end{figure}

\begin{figure}[tbh]
\begin{center}
\includegraphics[width=0.6 \hsize,clip]{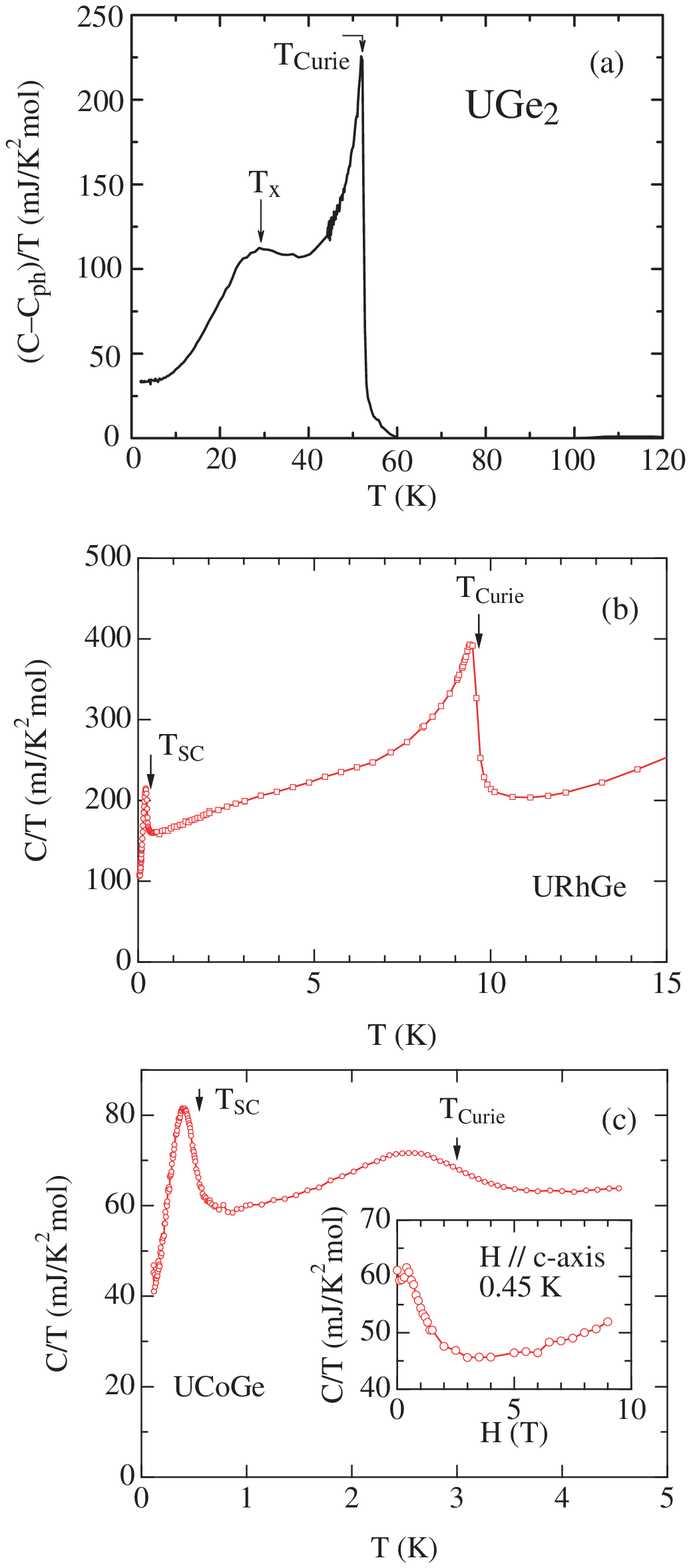}
\end{center}
\caption{Specific heat of UGe$_2$, URhGe and UCoGe. The phonon contribution is subtracted for UGe$_2$~\protect\cite{Har09_UGe2,Har_pub,Aok11_ICHE}.}
\label{fig:Cp}
\end{figure}

SC was discovered in the three uranium ferromagnets UGe$_2$, URhGe and UCoGe where a strong 5$f$ electronic component exists at the Fermi energy in the density of states. 
Thus, the $5f$ electrons are strongly delocalized. 
Figure~\ref{fig:energy_scale} summarizes their main characteristic parameters.
For UGe$_2$, FM appears at a rather high temperature, $T_{\rm Curie}\sim 52\,{\rm K}$, 
which is quite comparable to the renormalized bandwidth $W$.
The existence of the Fermi surface will be felt below $T\sim W/10$.
At ambient pressure the specific heat exhibits a clear jump at $T_{\rm Curie}$, as shown in Fig.\ref{fig:Cp}(a)~\cite{Har09_UGe2}. 
At $T_{\rm x}\sim 25\,{\rm K}$,
a crossover occurs between two interfering FM phases; FM2 with a sublattice magnetization $M_0$ $\approx$ $1.5\,\mu_{\rm B}$ and FM1 with $M_0\approx 1\,\mu_{\rm B}$~\cite{Pfl02}. 
SC appears only under pressure with a maximum $T_{\rm sc}^{\rm max}\sim 0.7\,{\rm K}$, at the pressure $P_{\rm x}\sim 1.2\,{\rm GPa}$, where the ground state switches from FM2 to FM1~\cite{Hux01}. 
A clear specific-heat anomaly was detected at T$_{x}(P)$ for $P\sim P_{\rm x}$~\cite{Tat04}. 
Finally, FM disappears at $P_{\rm c}\sim 1.49\,{\rm GPa}$.

URhGe with $T_{\rm Curie}=9.5\,{\rm K}$ is of great experimental interest because it becomes superconducting with $T_{\rm sc}=0.27\,{\rm K}$ at ambient pressure (Fig.~\ref{fig:Cp}(b)). 
Thus, it offers the possibility of applying a larger variety of experimental methods for understanding its superconducting properties. 
$T_{\rm Curie}$ appears well below the characteristic temperature related to the bandwidth ($W$). 
The low-temperature Sommerfeld coefficient is equal to $160\,{\rm mJ/K^2 mol}$~\cite{Aok11_ICHE,Har_pub}. 
With increasing pressure, $T_{\rm Curie}$ increases while $T_{\rm SC}$ decreases~\cite{Har05_pressure}. 
Thus, URhGe can be considered as a good example of the interplay between SC and FM which is far from the critical regime around $P_{\rm c}$.

Again, $T_{\rm Curie}$ is much lower than $W$ in UCoGe. 
However, the specific heat anomaly at $T_{\rm Curie}\sim 3\,{\rm K}$, shown in Fig.~\ref{fig:Cp}, is very broad and highly dependent on the sample purity. 
In fact, there is evidence from NMR measurements that the transition is indeed first order~\cite{Oht10} and the specific-heat anomaly results from the discontinuous change in entropy at $T_{\rm Curie}$ and strong fluctuations, which indicate that the magnetic coherence length remains constant over an extended $T$ window.
UCoGe is an unique example of ferromagnetic superconductivity at ambient pressure with a rather small ordered moment $M_0\sim 0.05\,\mu_{\rm B}$. 
With increasing pressure, $T_{\rm Curie}$ decreases and vanishes at $P_{\rm c}$ while $T_{\rm sc}$ initially raises and exhibits a broad maximum at $P_{\rm c}$~\cite{Has08_UCoGe,Has10,Slo09}.
Another unique feature of UCoGe is its low carrier concentration which implies that the contribution of the Co $3d$ states to the density of states is not negligible at the Fermi energy~\cite{Sam10,Pro10}.

\begin{figure}[tbh]
\begin{center}
\includegraphics[width=0.4 \hsize,clip]{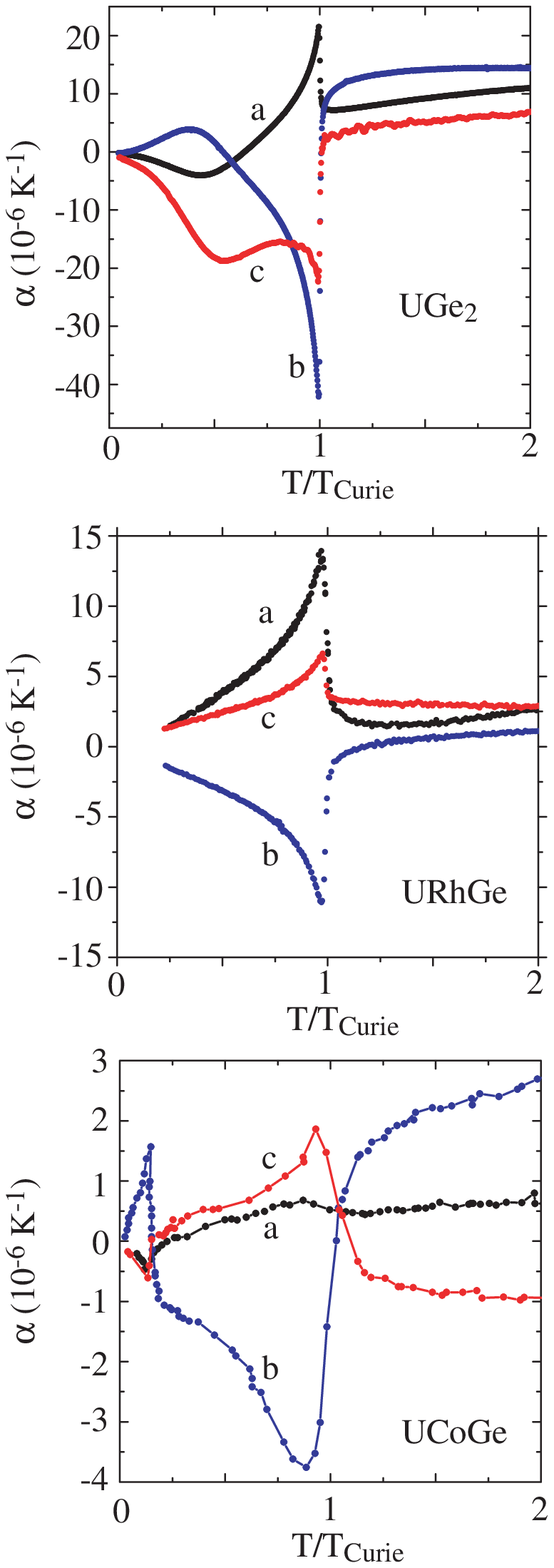}
\end{center}
\caption{Temperature dependence of the thermal-expansion coefficients along $a$, $b$ and $c$-axis in UGe$_2$, URhGe and UCoGe~\protect\cite{Har09_UGe2,Har_pub,Gas10}.}
\label{fig:Texp}
\end{figure}

As these compounds all have an orthorhombic structure, it is interesting to study their thermal-expansion coefficients along the three principal axes $a$, $b$ and $c$,~\cite{Har09_UGe2,Har_pub,Gas10,Aok11_ICHE} which are related to the uniaxial pressure derivative of $T_{\rm Curie}$ via the Ehrenfest and Clausius-Clapeyron relations for second-order and first-order transitions, respectively. 
As illustrated in Fig.~\ref{fig:Texp}, the thermal-expansion coefficients along the three axes do not show the same variation with uniaxial pressure at $T_{\rm Curie}$. 
For these three compounds, a large negative drop of $\alpha_b$ is observed at $T_{\rm Curie}$. 
It must be noted that for UGe$_2$, the crossover regime at $T_{\rm x} \sim 25\,{\rm K}$ from FM1 to FM2 is marked by extrema in $\alpha_a$ $\alpha_b$ and $\alpha_c$ which do not coincide in position. Above the critical pressure $P_{\rm c}$, 
where the system switches from FM1 to FM2 through a real first order transition, the jumps measured along the three axis have to occur at the same temperature~\cite{Kab09}. 
In UCoGe, the thermal expansion was also measured below $T_{\rm sc}$. 
The volume changes at $T_{\rm Curie}$ and $T_{\rm sc}$ are opposite in sign,
as observed in other highly anisotropic materials like URu$_2$Si$_2$ where the volume changes at the hidden order transition and $T_{\rm sc}$
are opposite in sign, as well.

The ratio of the volume thermal-expansion coefficient to the specific heat gives the opportunity to calculate the electronic Gr\"{u}neisen parameter $\Omega_e(T)$. 
Above $T_{\rm Curie}$, the three compounds have a positive Gr\"{u}neisen coefficient: 
the pressure derivative of the entropy, $dS/dP$ is negative. 
For URhGe, this sign remains the same on cooling through $T_{\rm Curie}$ since $T_{\rm Curie}$ increases with pressure. 
However, a sharp sign change occurs for UCoGe and UGe$_2$ ($dS/dP$ becomes positive) in excellent agreement with the observation that 
$T_{\rm Curie}$ collapses at $1\,{\rm GPa}$ and $1.5\,{\rm GPa}$ in UCoGe and UGe$_2$, respectively. 
In UGe$_2$, it is interesting to remark that $T_{\rm Curie}$ is comparable to $W$ and that the electronic Gr\"{u}neisen parameter in the PM phase is quite close to zero. For UCoGe, $\Omega_e(T)$ is already large and temperature independent above $T_{\rm Curie}$ with a value quite similar to that of the intermediate valence Ce compounds.
\begin{figure}[tbh]
\begin{center}
\includegraphics[width=0.6 \hsize,clip]{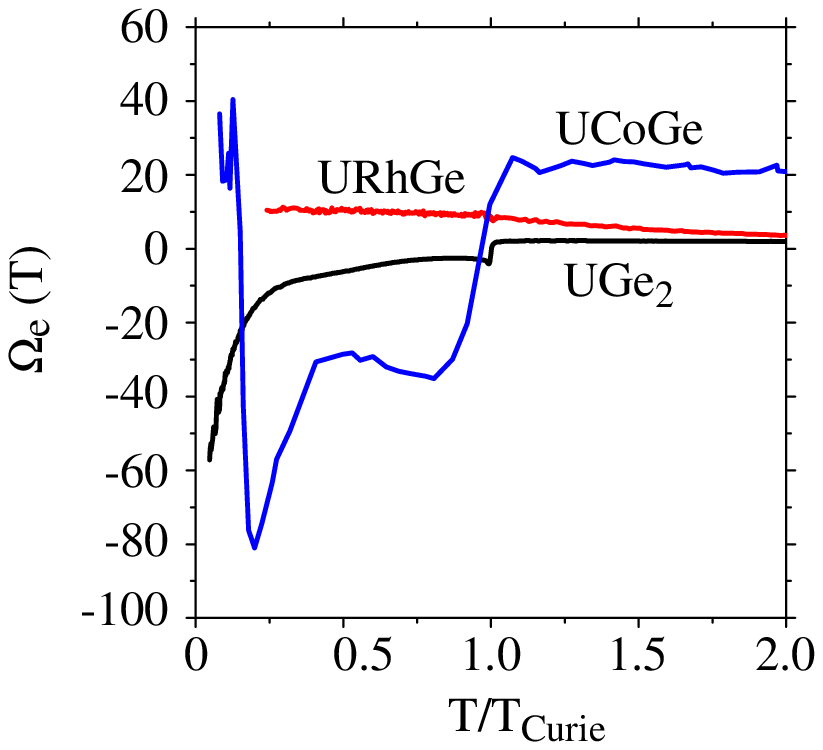}
\end{center}
\caption{Temperature dependence of the Gr\"{u}neisen parameter in UGe$_2$, URhGe and UCoGe~\protect\cite{Har_pub}.}
\label{fig:Gruneisen}
\end{figure}

\section{($T,P$) phase diagram: interplay of SC, PM and FM}
\begin{figure}[tbh]
\begin{center}
\includegraphics[width=1 \hsize,clip]{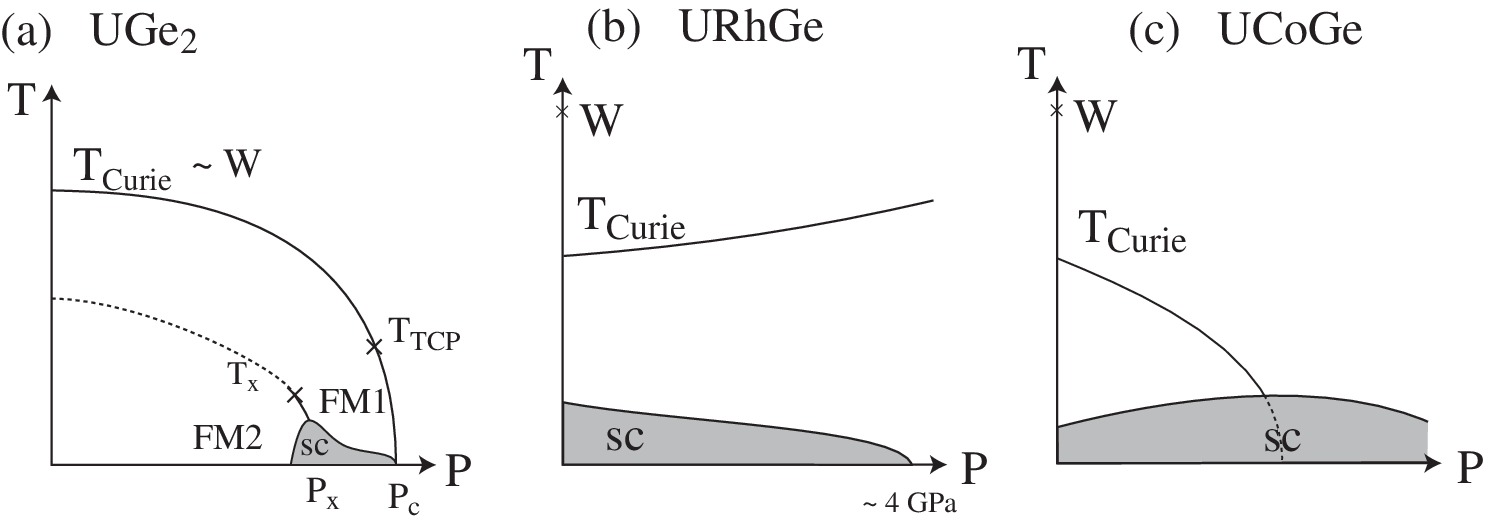}
\end{center}
\caption{Schematic ($T,P$) phase diagrams of UGe$_2$, URhGe and UCoGe.}
\label{fig:TP}
\end{figure}

Figure~\ref{fig:TP} shows schematic ($T,P$) phase diagrams of UGe$_2$, URhGe and UCoGe. 
In UGe$_2$, SC is squeezed between the two first-order transitions at $P_{\rm x}$ and $P_{\rm c}$~\cite{Sax00,Hux01,Ban07}. 
The robust first-order nature of these transitions makes it difficult to establish
whether SC exists homogeneously in the FM2 and PM phases and 
a definite conclusion is still under debate. 
Furthermore, the Fermi surface changes between the FM2 and the PM states~\cite{Set02,Ter01}. 
Two different models were proposed to explain the maximum of $T_{\rm sc}$ at $P_{\rm x}$. In the first one, SC is mediated by the charge density wave or spin density wave (CDW/SDW) fluctuations at $P_{\rm x}$~\cite{Wat02} while the second one invokes a twin-peak structure in the electronic density of states~\cite{San03}. 
No extra superstructures were observed at $P_{\rm x}$. 
The transition from FM2 to FM1 seems restricted to a switch between two FM states with consequences on $\lambda_z$ and $\lambda_{\Delta}$ reproducing rather well the pressure variation of $T_{\rm sc}$.

In URhGe, the situation corresponds to the behavior predicted in the FM domain for $P \ll P_{\rm c}$ with
the particularity that $T_{\rm Curie}$ increases with $P$ while $T_{\rm sc}$ decreases and disappears above $4\,{\rm GPa}$ (Fig.~\ref{fig:TP}(b))~\cite{Har05,Miy09}.

In UCoGe, $T_{\rm Curie}$ and $T_{\rm sc}$ tend to merge with increasing pressure~\cite{Has08_UCoGe,Slo09} and the FM anomaly is no longer detected in resistivity and susceptibility measurements when $T_{\rm sc}$ $\approx$ $T_{\rm Curie}$. Thus, $T_{\rm Curie}$ seems to collapse suddenly under pressure leaving a wide maximum in the pressure dependence of $T_{\rm sc}$. At least, the observation of the SC anomaly in the PM side indicates that bulk superconductivity exists in the PM domain~\cite{Has10}.

\section{Longitudinal and transverse magnetic field response}
In these Ising ferromagnets, the magnetic field leads to a particular response when it is applied 
either parallel or perpendicular to the initial sublattice magnetization $M_0$ (oriented along the $a$ axis 
for UGe$_2$ and along the $c$ axis for URhGe and UCoGe).

In UGe$_2$, the FM transition at $T_{\rm Curie}$ at ambient pressure is of second order and the application of a magnetic field parallel to $M_0$ weakens the FM correlations. Thus, the FM specific-heat anomaly is rapidly reduced and shifts to higher temperature with increasing $H$; $T_{\rm Curie}$ seems to increase with $H$ but $\gamma$ reaches the band-mass value $\gamma_{\rm B}$ when the field strength is comparable to the molecular field. In UGe$_2$, this molecular field is very large $\sim 200\,{\rm T}$.

\begin{figure}[tbh]
\begin{center}
\includegraphics[width=0.8 \hsize,clip]{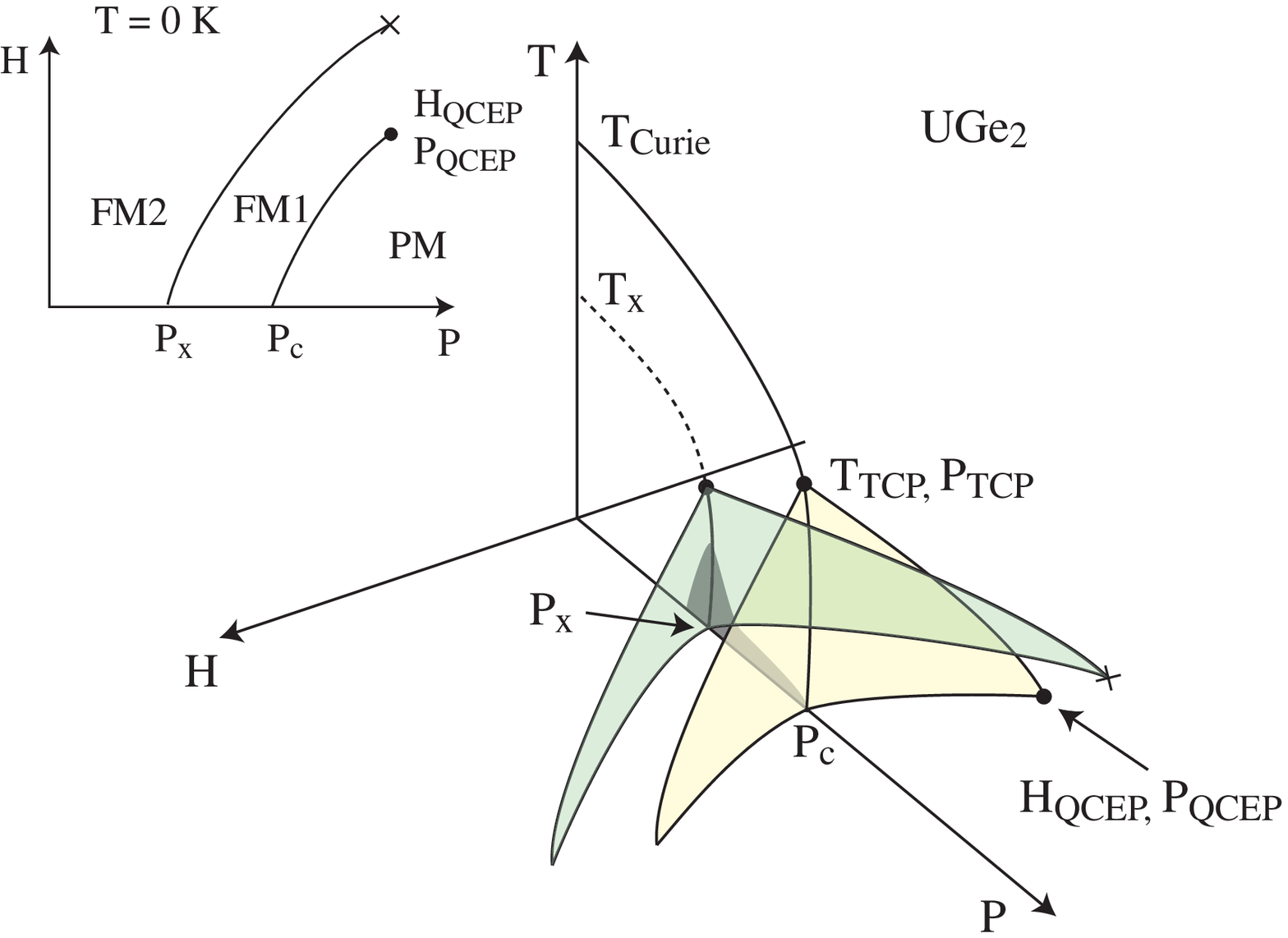}
\end{center}
\caption{Schematic ($T,P,H$) phase diagram of UGe$_2$. The inset shows the ($H,P$) plane at $T=0$.
Very recent experiments allow us to locate the position of the QCEP for $H_{\rm c}$.
The QCEP for $H_x$ requires a new set of high field measurements.}
\label{fig:UGe2_TPH}
\end{figure}

\begin{figure}[tbh]
\begin{center}
\includegraphics[width=0.6 \hsize,clip]{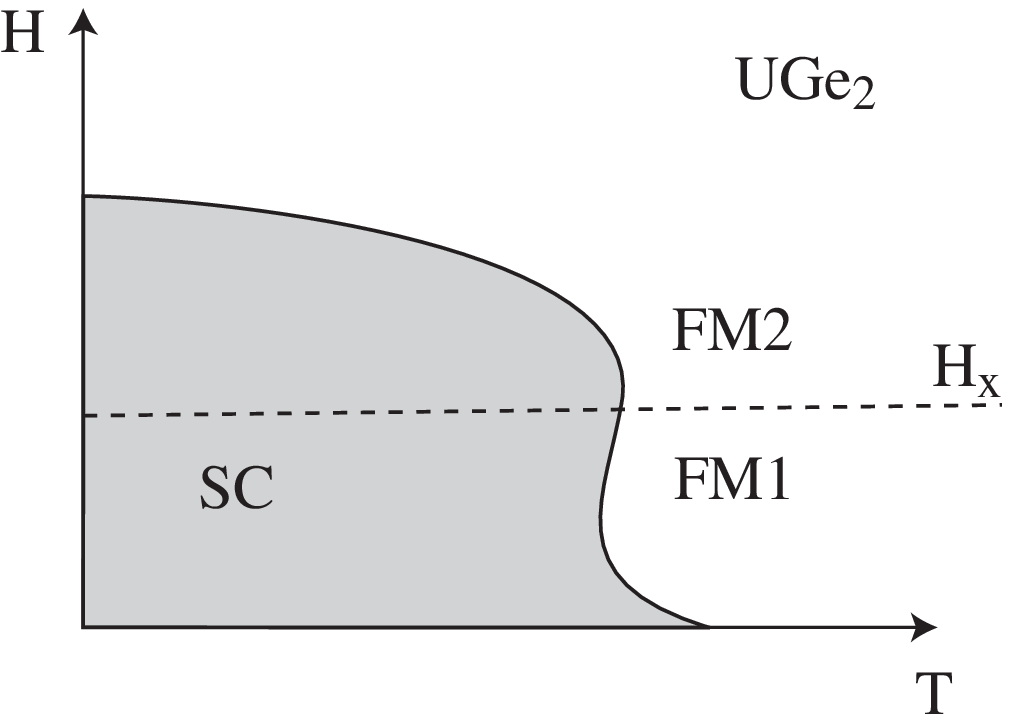}
\end{center}
\caption{Schematic temperature dependence of the upper critical field $H_{\rm c2}$ of UGe$_2$ for $H\parallel M_0$ ($a$-axis).}
\label{fig:UGe2_Hc2}
\end{figure}

However, the nature of the transition at $T_{\rm Curie}$ changes under pressure from second to first order at the tricritical point ($T_{\rm TCP}$,$P_{\rm TCP}$)~\cite{Tau10,Kot11}.
When the field is applied along the sublattice magnetization $M_0$, 
the occurrence of this tricriticality gives rise to in-field FM wings that open 
at the TCP and terminate at quantum critical end-points located at ($P_{\rm QCEP}$, $H_{\rm QCEP}$) for $T=0$ (see Fig.~\ref{fig:UGe2_TPH}). 
The pressure difference ${P}_{\rm QCEP}-P_{\rm c}$ is related to the pressure difference $\tilde{P}_{\rm c}-P_{\rm c}$ which correlates with the jump $\Delta M_0$ observed at $P_{\rm c}$. 
UGe$_2$ represents an ideal case for studying FM tricriticality since $\Delta M_0 \sim 0.9\,\mu_{\rm B}$ is large and the TCP ($T_{\rm TCP}=24\,{\rm K}$,$P_{\rm TCP}=1.42\,{\rm GPa}$) and the QCEP ($P_{\rm QCEP} \approx 3.5\,{\rm GPa}$,$H_{\rm QCEP}\approx 18\,{\rm T}$) are accessible with present laboratory equipments. 
Figure~\ref{fig:UGe2_TPH} shows the phase diagram of UGe$_2$ for $H\parallel M_0$. The ($T_{\rm x},P_{\rm x}$) line terminates at a critical point in the $H=0$ plane. 
UGe$_{2}$ switches from FM2 to FM1 at $H=H_x$ and from PM to FM1 at $H_{\rm c}$. 
Both fields $H_{\rm c}$ and $H_x$ will end at a QCEP.
High magnetic-field measurements, $H > 20\,{\rm T}$, are necessary to clarify the QCEP for $H_{\rm c}$. The transition from FM1 to FM2 that occurs at $H_x$ has a strong feedback on SC as illustrated by the unusual temperature dependence of $H_{\rm c2}(T)$ (see Fig.~\ref{fig:UGe2_Hc2})~\cite{She01}. At H=0, changes of the effective mass enhancement were observed at $P_{\rm x}$ and $P_{\rm c}$, respectively. 
Thus, similar changes should also occur at $H_{\rm c}$ and $H_{\rm x}$ for $P_{\rm c}<P<P_{\rm QCEP}$;
on approaching $P_{\rm QCEP}$, $H_{\rm QCEP}$ well defined maximum of $m^\ast(H)$ must appear.
An interesting point is the possible concomitant occurrence of Lifshitz transition which is associated with the topological change of 
Fermi surfaces~\cite{Yam07_Lifshitz}.
This may add another feedback for the treatment of the metamagnetic transition.

For $H\parallel M_0$, a weakening of the FM correlations similar to that of UGe$_2$ at ambient pressure is observed in URhGe and UCoGe. In URhGe, the FM transition is second order at $P=0$ and moves away from tricriticality as $P$ is increased since $\partial T_{\rm Curie}/\partial P > 0$. For UCoGe, as already mentioned, the PM-FM transition may be first order~\cite{Oht08}. However, $M_0$ is already weak at ambient pressure and it decreases with $P$. It is thus suspected that $P_{\rm QCEP}$ will be very close to $P_{\rm c}$ and that $H_{\rm QCEP}$ will be rather low.

\begin{figure}[tbh]
\begin{center}
\includegraphics[width=0.8 \hsize,clip]{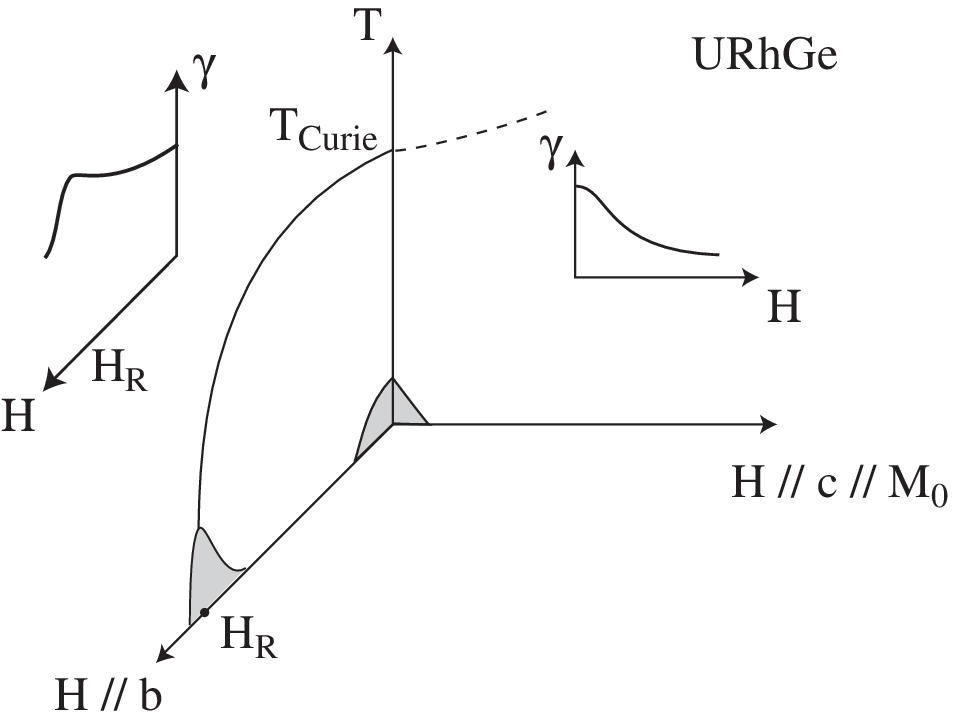}
\end{center}
\caption{Schematic $(T,H)$ phase diagram for $H\parallel b$ (hard-axis) and $H \parallel c$ (easy-axis). The field dependence of $\gamma$ is also depicted for both cases.}
\label{fig:URhGe_schematic}
\end{figure}

However, spectacular effects arise for $H\perp M_0$. The transverse response leads to a decrease of $T_{\rm Curie}$, which can be described using the Landau free energy~\cite{Min11}. 
Figure~\ref{fig:URhGe_schematic} shows schematically the field variation of $T_{\rm Curie}(H)$ and $\gamma(H)$ for $H\parallel M_0$ and $H\perp M_0$ in URhGe. If $\gamma$ goes through a maximum, a field enhancement of $m^{\ast\ast}$ accompanied by an enhancement of $T_{\rm sc}$ occurs when the induced transverse magnetic component along the hard axis, {\it e.g.} $\chi_{b}H_b$ along $b$-axis, becomes comparable to $M_0$ (where $\chi_b$ is the initial slope of magnetization along $b$ axis). Table~\ref{tab:UGe2_URhGe_UCoGe} gives the estimated characteristic fields along the three axes for the three uranium SC ferromagnets.
\begin{table}[tbhp]
\caption{Susceptibilities and characteristic fields of UGe$_2$, URhGe and UCoGe.}
\begin{center}
\begin{tabular}{ccccccc}
\hline
			& $\chi_a$	& $\chi_b$	& $\chi_c$			& $H_a$	& $H_b$	& $H_c$		\\
			& \multicolumn{3}{c}{($\mu_{\rm B}/{\rm T}$)}	& \multicolumn{3}{c}{(T)}		\\
\hline
UGe$_2$		& 0.006		& 0.0055		& 0.011				& 230		& 250		& 122		\\
URhGe		& 0.006		& 0.03		& 0.01					& 66		& 13		& 40			\\
UCoGe		& 0.0024	& 0.006		& 0.029				& 29		& 12		& 2.5		\\
\hline
\end{tabular}
\end{center}
\label{tab:UGe2_URhGe_UCoGe}
\end{table}%

\section{Reinforcement of SC in the transverse response}
\begin{figure}[tbh]
\begin{center}
\includegraphics[width=0.8 \hsize,clip]{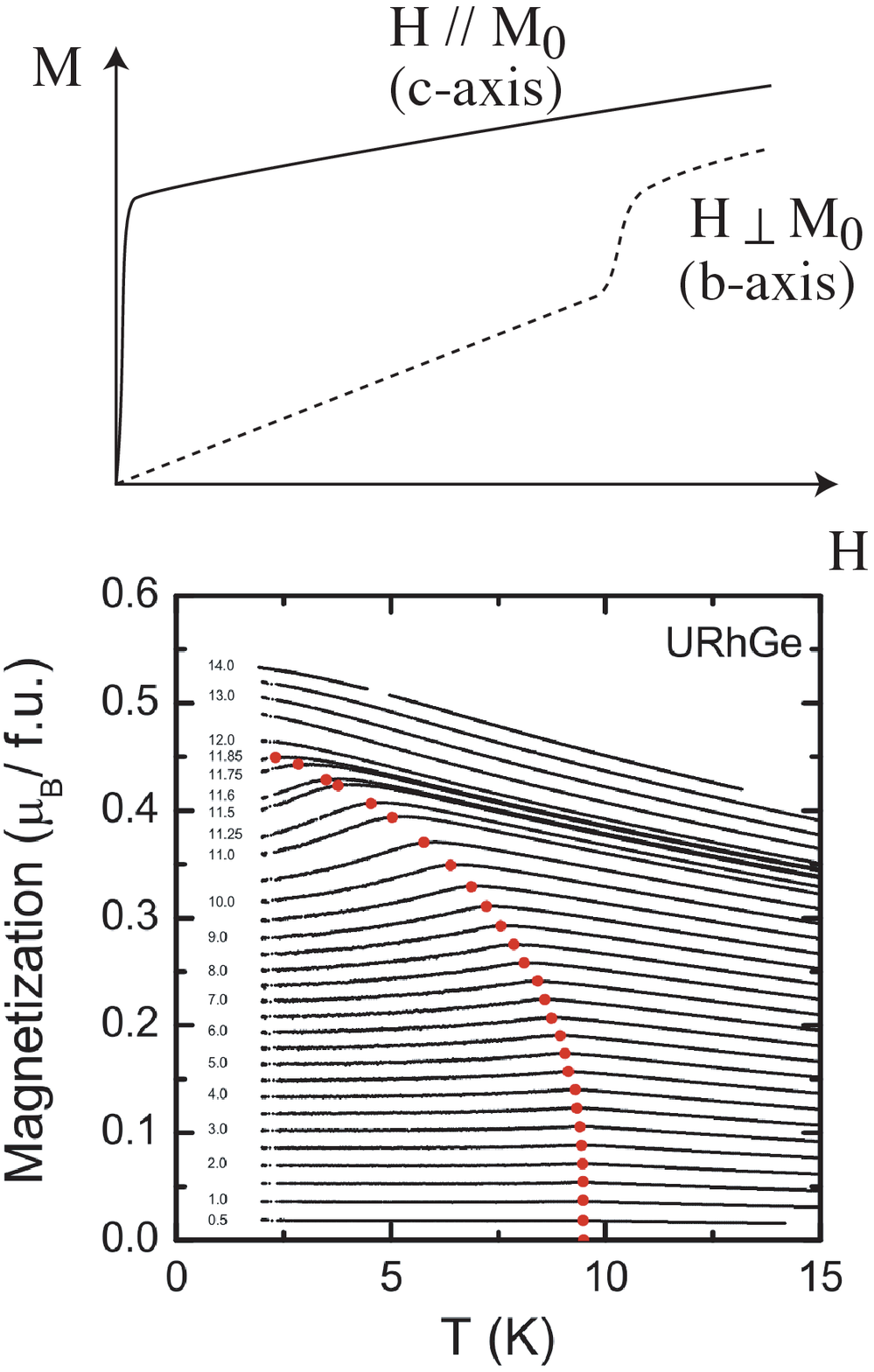}
\end{center}
\caption{Schematic magnetization curves at low temperature and the temperature dependence of magnetization at constant fields $H\parallel b$-axis in URhGe~\protect\cite{Har_pub}.}
\label{fig:URhGe_mag}
\end{figure}

\begin{figure}[tbh]
\begin{center}
\includegraphics[width=0.8 \hsize,clip]{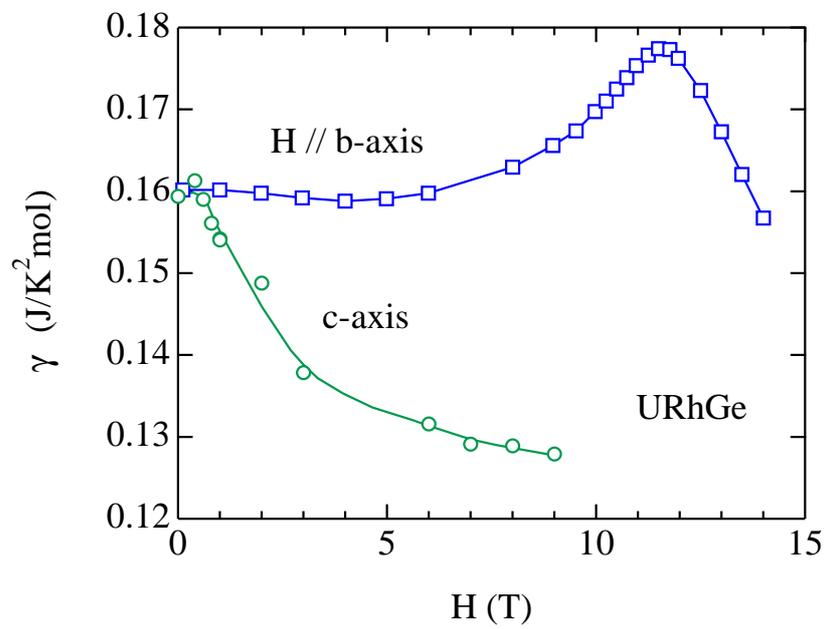}
\end{center}
\caption{Field dependence of the Sommerfeld coefficient $\gamma$ obtained from the Maxwell relation ($b$-axis) and direct specific-heat measurements at $0.4\,{\rm K}$ ($c$-axis) in URhGe~\protect\cite{Har_pub,Aok11_ICHE}.}
\label{fig:URhGe_gamma}
\end{figure}

\begin{figure}[tbh]
\begin{center}
\includegraphics[width=0.8 \hsize,clip]{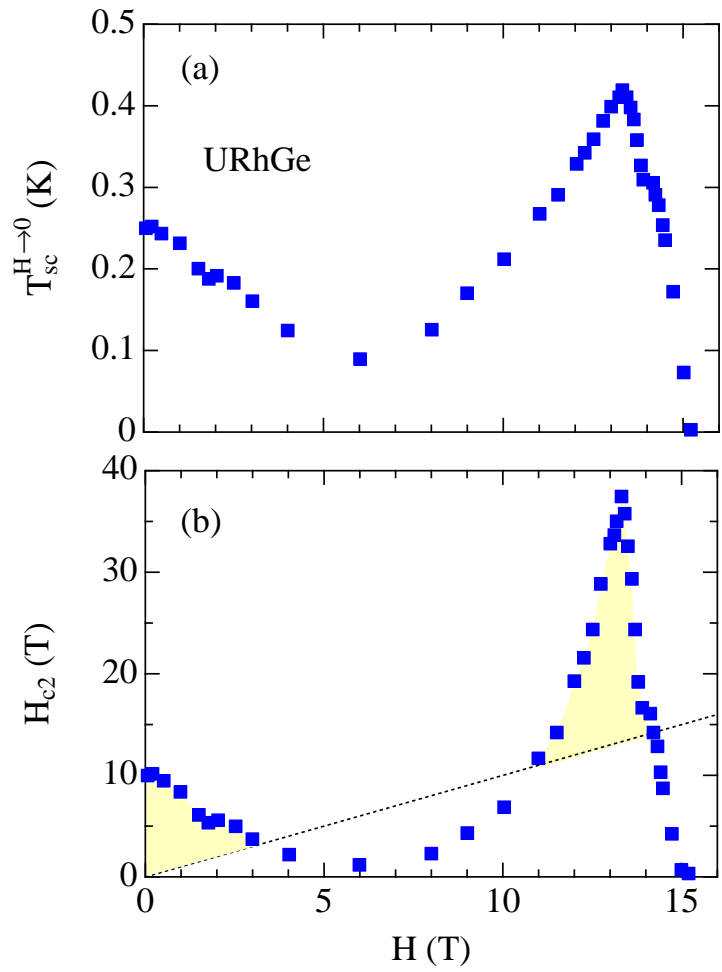}
\end{center}
\caption{Calculated $T_{\rm sc}$ and $H_{\rm c2}$ based on the field dependence of $m^\ast$ in URhGe~\protect\cite{Miy08}.}
\label{fig:URhGe_calc_Tc_Hc2}
\end{figure}

\begin{figure}[tbh]
\begin{center}
\includegraphics[width=1 \hsize,clip]{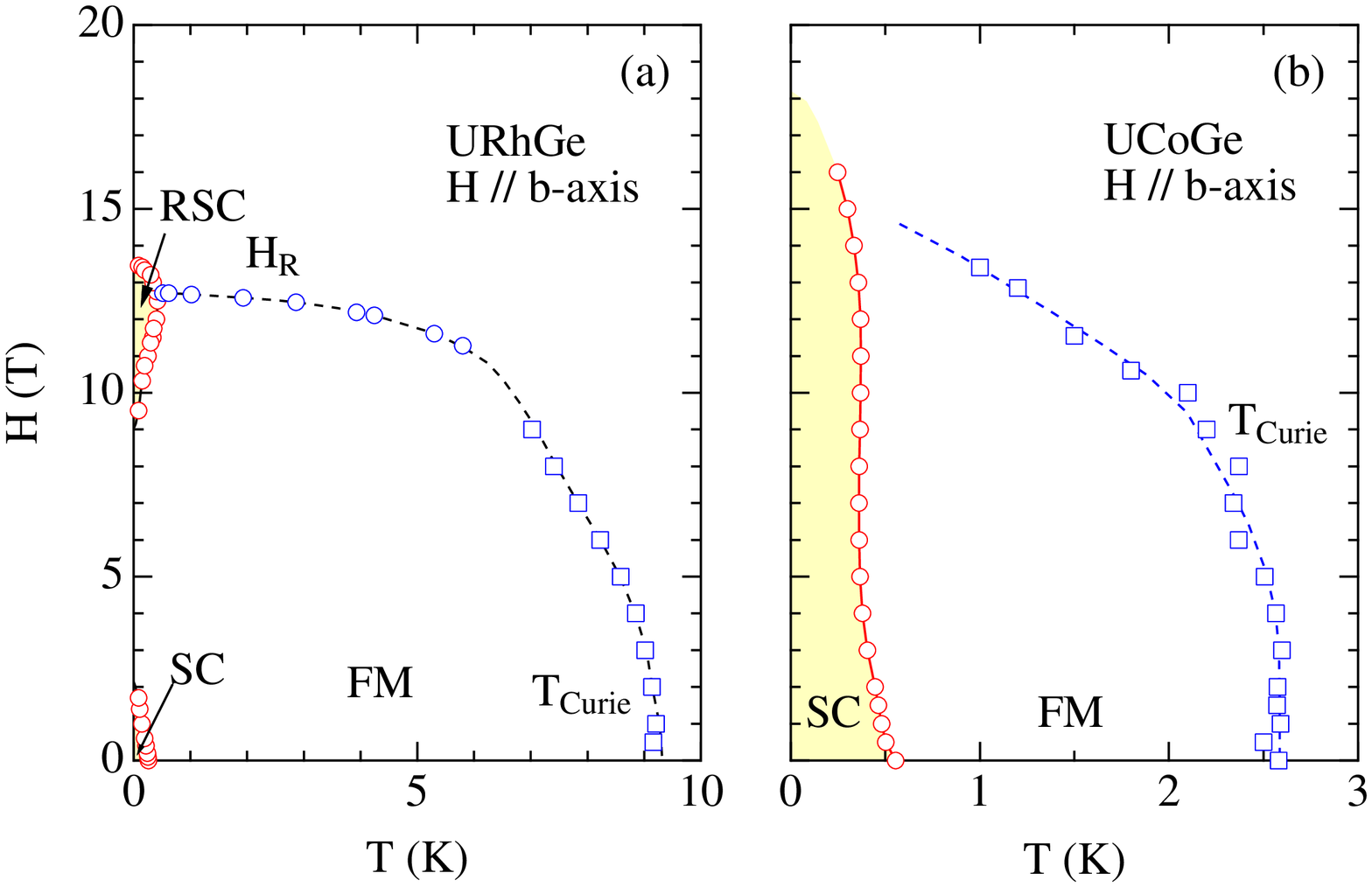}
\end{center}
\caption{$(H,T)$ phase diagrams of URhGe and UCoGe for $H\parallel b$-axis~\protect\cite{Aok11_ICHE,Aok09_UCoGe}.}
\label{fig:URhGe_UCoGe_HT_phase}
\end{figure}

In URhGe, the susceptibility along the hard magnetization axis $b$, $\chi_{\rm b}=\partial M_b/\partial H$, is large in comparison with the easy axis $c$, ($\chi_b/\chi_c \sim 3$). 
At a field $H_{\rm R}\parallel b$, a reorientation of the magnetic moment occurs and the easy axis changes from the $c$ to the $b$ axis. 
In a restricted field range centered around $H_{\rm R}$, reentrant SC appears. 
Figure~\ref{fig:URhGe_mag} shows schematic magnetization curves and the temperature dependence of $M$ at different fields $H\parallel b$-axis. 
$T_{\rm Curie}$ is marked by a maximum of $\chi_b$.
The coefficient of the magnetization $T^2$ term is linked to the Sommerfeld coefficient, 
according to the thermodynamic Maxwell relation $\partial \gamma/ \partial H = \partial^2 M / \partial T^2$. 
As shown here, $T_{\rm Curie}$ decreases with increasing field and is suppressed at $H_{\rm R}$ at low temperatures. 
The field dependence of the effective mass $m^\ast (H)$ obtained from the Maxwell relation and direct specific-heat measurements are shown in Fig.~\ref{fig:URhGe_gamma}. 
The enhancement of the effective mass with increasing field $H\parallel b$ is at the origin of the reentrant SC (RSC) illustrated in Fig.~\ref{fig:URhGe_calc_Tc_Hc2}.
Using the simple formula, $T_{\rm sc} \sim T_0 \exp (-m^\ast/m^{\ast\ast} )$, $H_{\rm c2}$ can be calculated within the orbital limit: $H_{\rm c2}\sim (m^\ast T_{\rm sc})^2$. 
Excellent agreement is obtained for the magnetic field range where RSC is observed (Fig.~\ref{fig:URhGe_UCoGe_HT_phase}). Moreover, knowing the $P$ dependence of $m^\ast(H)$, RSC is predicted to collapse at $P_{\rm RSC}\sim 2\,{\rm GPa}$ as observed experimentally~\cite{Miy09}.
At $0\,{\rm K}$, we notice that a linear extrapolation of M(H), for $H\parallel b$, from $H > H_{R}$ to
$H=0$ exhibits a non-zero intercept, suggesting that the reorientation process does not correspond to a transition to the PM regime. 
The preservation of the FM phase suggests that the FM Fermi surface is rather robust during the reorientation process, 
in good agreement with the weak singularities of the thermoelectric power detected at $H_{\rm R}$~\cite{Mal_pub}.

\begin{figure}[tbh]
\begin{center}
\includegraphics[width=0.8 \hsize,clip]{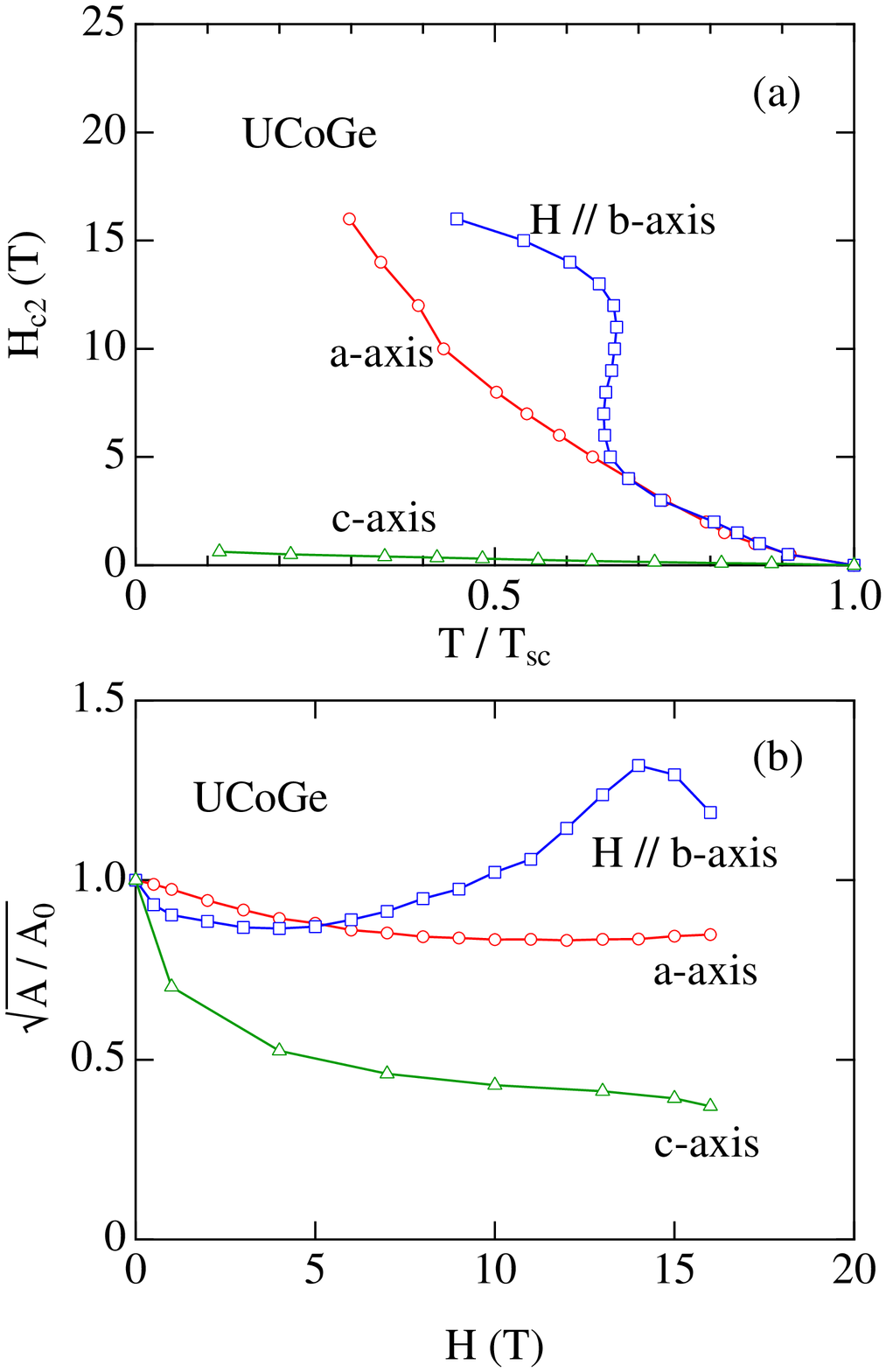}
\end{center}
\caption{(a) Temperature dependence of $H_{\rm c2}$ along the orthorhombic directions of UCoGe (b) Field dependence of effective mass (the normalized $\sqrt{A}$)~\protect\cite{Aok09_UCoGe}.}
\label{fig:UCoGe_Hc2_A_coef}
\end{figure}

In UCoGe, for the same field strength, no reorientation is expected since $\chi_c$ is larger than $\chi_b$ and $\chi_a$. 
However, the transverse response, when $H$ reaches $H_b$, leads to an unusual dependence of $H_{\rm c2}(T)$ (as shown in Fig.~\ref{fig:URhGe_UCoGe_HT_phase}). 
It is related to a field enhancement of $m^\ast$ as reflected by the enhancement of $A(H)$ for $H\parallel b$ when $H$ approaches $H_b$. 
For $H\parallel c$ a strong decrease of $A$ is detected (Fig.~\ref{fig:UCoGe_Hc2_A_coef}). 
The calculated Fermi surface of UCoGe in the FM phase is quite different from that in the PM phase,~\cite{Sam10} and the system is close to a FM--PM instability. 
Hence, it could be possible that the transverse magnetic field drives the system through the FM--PM singularity. 
Evidence could be given by the recent observation of large variations of the thermoelectric power at $H_b$~\cite{Mal_pub}. 
In recent Shubnikov-de Haas experiments that measure the Fermi surface and the cyclotron effective mass, a quite large $H$ response is detected~\cite{Aok11_UCoGe}.

\section{Conclusion and remarks}
We have presented the ($T,P,H$) phase diagrams of the three uranium ferromagnetic superconductors, UGe$_2$, URhGe and UCoGe. 
We have focused on the enhancement of effective mass and its relation to SC. 
We emphasize that the magnetic singularities at $P_{\rm c}$, $P_{\rm x}$, $H_x$, and $H_{\rm c}$ are often associated with a Fermi surface reconstruction related to the first-order nature of the magnetic transition in UGe$_2$ and UCoGe. 
The full determination of the Fermi surface, as a function of $P$ and $H$, is expected soon thanks to progresses in the crystal purity. 
The case of URhGe which is located far from FM--PM instability and far from tricriticality seems to be the ideal example of FM superconductivity.

An interesting aspect of FM superconductivity concerns the influence of FM on macroscopic phenomena such as the Meissner effect~\cite{Sho05}. 
Other topics are SC in FM domain walls and the phenomena associated with the relative orientation of the SC order parameter to the magnetization, 
the effect of SC on FM domain structure~\cite{Buz03}. Of course as the internal field is large with respect to the lower critical field $H_{\rm c1}$ ($\sim 10^{-3}\,{\rm T}$),
spontaneous vortex formation may already occur at $H=0$. It is only recently that careful DC magnetization measurement were realized in the case of UCoGe ($4\pi M\sim 0.01\,{\rm T}$), no full Meissner effect, 
i.e. no indication of $H_{\rm c1}$ was detected at least for $H \parallel c$-axis.~\cite{Deg10,Pau_pub}.

Advances in the field have been mainly achieved through the discovery of new systems. 
Even now the main goal is to discover a very clean system like Ce-115 heavy fermion superconductors where large and pure single crystals are easily available. 
Unfortunately up to now, high quality single crystal growth of UCoGe and URhGe is a difficult task and for UGe$_2$ SC appears only under pressure squeezed between two first order transitions.
It is worthwhile to remark that SC in FM materials has only been detected in uranium intermetallic compounds with Ising type FM, confirming the key role of longitudinal fluctuation and for materials with quite moderate heavy fermion character ($\gamma \le 160\,{\rm mJ/K^2 mol}$) by comparison to the large value of $\gamma$ ($>1\,{\rm J/K^2 mol}$) reported for prototype $d$-wave superconductors (CeCu$_2$Si$_2$, Ce-115) close
to their AF--PM instability.
Maybe due to the low value of $T_{\rm sc}$
provided by FM longitudinal fluctuations and the sensitivity to disorder at both FM and SC onsets,
a moderate renormalized band width is a quite favorable condition for the coexistence of FM and SC.
Up to now, all attempts to discover SC in other FM materials have failed with Ce ferromagnetic
heavy fermion compounds.

\section*{Acknowledgments}
We thank J. P. Brison, A. Buzdin, S. Fujimoto, H. Harima, K. Hasselbach, E. Hassinger, L. Howald, K. Ishida, W. Knafo, G. Knebel, H. Kotegawa, L. Malone, C. Meingast, V. Michal, V. Mineev, K. Miyake, C. Paulsen, S. Raymond, R. Settai, I. Sheikin and Y. Tada for fruitful discussion. 
This work was supported by ERC starting grant (NewHeavyFermion) and French ANR project (CORMAT, SINUS, DELICE).
J. F. is supported as a ``directeur de recherche \'{e}m\'{e}rite'' in CNRS.

\end{document}